# An Enabling Waveform for 5G - QAM-FBMC: Initial Analysis

Yinan Qi and Mohammed Al-Imari

Samsung Electronics R&D Institute UK, Staines-upon-Thames, Middlesex TW18 4QE, UK

{yinan.qi, m.al-imari}@samsung.com

*Abstract*— In this paper, we identified the challenges and requirements for the waveform design of the fifth generation mobile communication networks (5G) and compared Orthogonal frequency-division multiplexing (OFDM) based waveforms with Filter Bank Multicarrier (FBMC) based ones. Recently it has been shown that Quadrature-Amplitude Modulation (QAM) transmission and reception can be enabled in FBMC by using multiple prototype filters, resulting in a new waveform: QAM-FBMC. Here, the transceiver architecture and signal model of QAM-FBMC are presented and channel estimation error and RF impairment, e.g., phase noise, are modeled. In addition, initial evaluation is made in terms of out-of-band (OOB) emission and complexity. The simulation results show that QAM-FBCM can achieve the same BER performance as cyclic-prefix (CP) OFDM without spectrum efficiency reduction due to the adding of CP. Different equalization schemes are evaluated and the effect of channel estimation error is investigated. Moreover, effects of the phase noise are evaluated and QAM-FBMC is shown to be robust to the phase noise.

## I. INTRODUCTION

With the 5G research well underway in many parts of the world, researchers are looking at new waveforms to fulfill the stringent requirements imposed by 5G services [1]. The waveform design for 5G should address the challenges imposed by extremely diverse use cases, deployment scenarios and service requirements, including many aspects.

Firstly, high spectral efficiency needs to be achieved. Current multicarrier technique, i.e. OFDM, that is adopted in 3GPP-LTE/LTE-A and 802.11.a/g/n uses cyclic prefix (CP) to eliminate inter-symbol interference when the delay spread of the channel is lower than the CP length. However, as it is redundant data, the use of CP leads to a loss of spectral efficiency. In addition, the employment of rectangular pulses in CP-OFDM leads to strong out-of-band (OOB) emission, and hence, guard bands are required. Thus, 5G waveforms are expected to have high spectral efficiency by avoiding the use of CP, and have low OOB emission to eliminate the need for guard bands.

Secondly, efficient use of non-contiguous spectrum should be taken into account. Considering that the availability of large amount of contiguous spectrum is getting very difficult, especially for below 6GHz bands, the aggregation of non-contiguous frequency bands is considered for 5G systems to make the best use of the scarce spectrum and meet the increasing demand for emerging wireless applications. The need for efficient utilization of the non-contiguous spectrum has motivated the search for multicarrier waveforms that provide lower OOB emission without sacrificing spectral efficiency.

Thirdly, the new waveform should be compatible with multi-antenna technologies. To meet the high data rate demand and the increasing number of connected devices, 5G systems is expected to deliver an efficient use of the spectrum by using MIMO technologies (such as massive MIMO). Hence, waveforms that efficiently support MIMO are very desirable in 5G systems. For above 6 GHz bands, due to the hostile propagation condition in millimeter wave (mm-wave) radio channels, e.g., severe path loss, vulnerability to blockage, etc., large antenna gains at both transmitter and receiver sides are required to overcome propagation losses. As per the Friis Law the single antenna aperture size reduces with the square of the carrier frequency and negatively impacts the amount of radio energy captured at the receiver. In this regard, very large scale antenna arrays are needed that enable highly directive transmit and receive beamforming.

Fourthly, sporadic access should be supported. Machine Type Communications (MTC) is one of the prominent use cases for 5G systems. In most of the cases, MTC traffic involves relatively small packets per connection. In addition, a typical characteristic of MTC is that the devices must be robust, simple (i.e. low-cost) and be able to operate on substantially long battery lives. In order to reduce power consumption, low-power devices require transmitting their data immediately after waking up with very low overhead and enter a dormant state directly after data transmission. Thus, as devices cannot be fully synchronized in this scenario, 5G waveforms need to be robust against timing and frequency offset to limit the amount of required signalling.

Finally, the new waveforms should be robust against more severe RF impairments. 5G is expected to utilize mm-wave bands to support Extreme Mobile BroadBand (xMBB) and immersive multimedia experience, as mm-wave bands provide 10 times more bandwidth than the 4G cellular-bands. However, waveforms for above 6 GHz bands should address some unique challenges such as robustness against RF impairments, e.g., phase noise, I/Q-imbalance, Power Amplifier (PA) nonlinearity etc. Generally, RF impairments become more severe in mm-wave bands, e.g., the phase-noise variance grows with the square of carrier frequency. As a result, robustness against RF impairments needs to be enhanced.

In addition to the main challenges list above, there are many other issues need to be taken into consideration when designing 5G waveforms, such as time localization, Peak-to-Average Power Ratio (PAPR), etc.

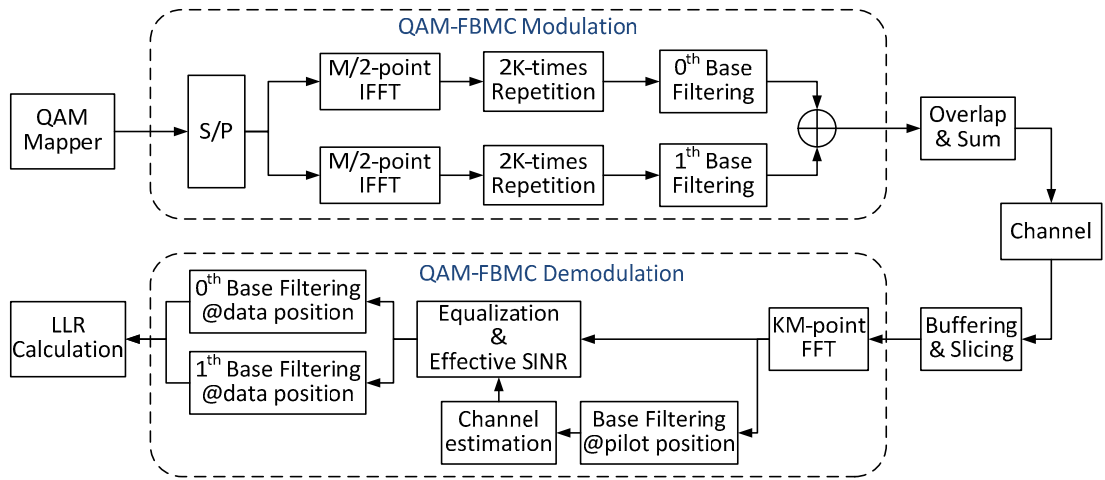

Fig. 1.     QAM-FBMC transceiver block diagram

The rest of the paper is organized at follows. In section II of this paper, a review and comparison between OFDM based waveforms and FBMC based waveforms are presented. Section III is devoted to the analysis of QAM-FBMC, including signal model, RF impairments, complexity analysis, etc. The early evaluation results are presented in section IV. The paper is concluded in section V, with some indications about the further work planned in this area.

## II. COMPARIOSN BETWEEN OFDM BASED AND FBMC BASED WAVEFORMS

### A. OFDM Based Waveforms

OFDM has been widely accepted in many wireless standards such as WiFi, WiMAX and LTE/LTE-A [2][3][4]. It is well-known to be able to provide simple transceiver design and easy integration with advanced multiple antenna technologies such as higher order MIMO and massive MIMO for beamforming. With the use of CP, multipath fading can be effectively tackled as long as the duration of CP is longer than the duration of channel impulse response. The drawbacks of CP-OFDM include comparatively high OOB leakage, posing the need to use large guard bands, and degraded overall spectral efficiency due to the added CP overhead, i.e., $TF<1$, with $T$ and $F$ representing the symbol interval and subcarrier spacing, respectively. The large side-lobe of CP-OFDM results in inter-carrier interference (ICI) when synchronization cannot be achieved or when Doppler Spread happens in the cases of high mobility users, degrading the overall system performance. Under such circumstances, enhanced OFDM based waveforms are being investigated currently to address the requirements for 5G wireless systems and to enable a more flexible adaptation to the needs of the diverse services.

The unfavourable spectrum confinement property of OFDM is due to the usage of rectangular pulses whose power spectrum density (PSD) in the frequency domain is a sinc function. Filtered-OFDM (F-OFDM), one the other hand, filters the subcarriers to achieve much smaller OOB leakage while at the same time maintaining strict separation of the signals in the time domain and the complex field orthogonality [5]. Universal Filtered OFDM (UF-OFDM), i.e., UFMC, filters a block of subcarriers, i.e., a sub-band to keep OOB emission low [6]. Nevertheless, all OFDM based waveforms including the enhanced waveforms cannot achieve maximum spectral efficiency, i.e., $TF=1$.

### B. FBMC Based Waveforms

FBMC, as an enabling technology for enhancing the fundamental spectral efficiency because of the well-localized time/frequency traits adopted from a pulse shaping filter per subcarrier, can reduce the overhead of guard band required to fit in the given spectrum bandwidth, while meeting the spectrum mask requirement [7]. These can be realized by expanding the prototype filter pulse and symbol duration over $L$ symbol intervals in the time domain, resulting in overlapping pulses with duration $LT$, where $L$ is referred to as overlapping factor.

In FBMC, the effectively increased symbol duration is suitable for handling the multipath fading channels even without CP overhead. Consequently, the FBMC system can reduce the inherent overheads such as CP and guard-bands in CP-OFDM. FBMC is also attractive in specific asynchronous scenarios, where Coordinated Multi-Point Transmission and Reception (CoMP) and Dynamic Spectrum Access (DSA) in a fragmented spectrum are employed to support the much higher traffic demand in 5G.

In order to achieve the maximum spectral efficiency, Offset QAM (OQAM) has been proposed in the literature. However, the orthogonality in OQAM is in the real signal field only. To maintain the transmission symbol rate, the conventional FBMC system generally doubles the lattice density either in time or in frequency compared with OFDM while adopting OQAM, where in-phase and quadrature-phase modulation symbols are mapped separately with half symbol duration offset. Since complex domain orthogonality cannot be achieved, OQAM-FBMC or staggered multitone (SMT) causes intrinsic interference that makes it not straightforward to apply conventional pilot designs and corresponding channel estimation algorithms as well as MIMO schemes as in CP-OFDM systems. In this regard, QAM-FBMC system which can transmit the QAM symbols is proposed to enable fundamental spectral efficiency enhancement whilst keeping the signal processing complexity low [8][9].

## III. QAM-FBMC

### A. System Model

The transceiver architecture of QAM-FBMC is shown in Fig. 1. The information symbols are divided into the even-numbered sub-carrier symbols and the odd-numbered sub-carrier symbols. Then the symbols are IFFT transformed and repeated. Finally pulse shaping with two prototype filters are performed to the symbols by means of windowing (element-wise multiplication) and added. The symbol interval of QAM-FBMC is the same as the duration of CP-less OFDM symbols but the symbols overlap with each other. The received symbols are FFT transformed and equalized in frequency domain. Then the received symbols are filtered by the Rx filter for even-numbered sub-carrier symbols and due to certain orthogonality conditions, the odd-numbered sub-carrier symbols are filtered out. By doing this, the received symbols are divided into the even-numbered symbols and the odd-numbered symbols. Each symbol is then demodulated.

### B. Signal Model

The QAM-FBMC system separates adjacent subcarriers with $B$ filter-banks to keep orthogonality in complex domain. The transmitted signal can be expressed as

$$x[n] = \sum_{k=-\infty}^{\infty} \sum_{b=0}^{B-1} p_{T,b,0}[n-kM] \left( \sum_{s=0}^{\frac{M}{B}-1} D_{b,s}[k] e^{j\frac{2\pi n}{M/B}s} \right), \quad (1)$$

where $D_m[k]$ is the complex data symbol on the $(sB+b)$-th subcarrier in the $k$-th symbol, $M$ is the total number of subcarriers and $p_{b,s}[n]$ is the $b$-th prototype filter. The prototype filters are defined as

$$p_{b,s}[n] \triangleq q_b[n] e^{j\frac{2\pi}{M}(Bs+b)n}. \quad (2)$$

It can be efficiently implemented using $M/B$-point inverse fast Fourier transform (IFFT), $BL$-times repetition, and time domain base filtering. Here $L$ is the overlapping factor.

The $k$-th transmitted symbol vector $\mathbf{x}[k]$ of length $N=LM$ can be represented as

$$\mathbf{x}[k] = \mathbf{W}_N^H \mathbf{P}_f \underline{\mathbf{D}}[k], \quad (3)$$

where $\mathbf{P}_f$ is the frequency-domain filter coefficients matrix with size $N$-by-$M$ in which the $(Bs+b+1)$-th column is given by $N$-point DFT of the sibling filter $p_{b,s}[n]$ in (2), and $\mathbf{W}_N^H$ is $N$-point IDFT matrix. $\underline{\mathbf{D}}[k]$ is the transmitted data symbol vector in the $k$-th QAM-FBMC symbol.

After $N$-point FFT, the 0-th received QAM-FBMC symbol in frequency domain, $\underline{\mathbf{X}}_R[0]$, is given by

$$\underline{\mathbf{X}}_R[0] = \sum_{k=-L}^{L-1} \overline{\mathbf{H}}[k] \mathbf{P}_T \underline{\mathbf{D}}[k] + \underline{\mathbf{w}}[0], \quad (4)$$

where $\overline{\mathbf{H}}[k] = \mathbf{W}_N \mathbf{T}[k] \mathbf{H}[k] \mathbf{W}_N^H$ and $\underline{\mathbf{w}}[0]$ is the noise vector. Here the channel matrix $\mathbf{H}[k]$ is a Toeplitz matrix of size $(N+M)$-by-$N$ whose $n$-th column is given by the circular shift of the channel impulse response vector. The matrix $\mathbf{T}[k]$ of size $N$-by-$(N+M)$ represents the shift-and-slice procedure in Fig. 1, which captures the effect of interference to the 0-th symbol from the preceding and succeeding symbols, and it is defined as

$$\mathbf{T}[k] \triangleq \begin{cases} \begin{bmatrix} \mathbf{0} & \mathbf{I}_{N+M+kM} \\ \mathbf{0} & \mathbf{0} \end{bmatrix}, & k < 0 \\ [\mathbf{I}_N \quad \mathbf{0}], & k = 0 \\ \begin{bmatrix} \mathbf{0} & \mathbf{0} \\ \mathbf{I}_{N-kM} & \mathbf{0} \end{bmatrix}, & k > 0 \end{cases} \quad (5)$$

It should be noted that due to overlapping and sum transmit structure, interference from adjacent $(2L-1)$ transmit FBMC symbols are included in the received signal. Therefore, the matrix $\overline{\mathbf{H}}[k]$ is not diagonal and the off-diagonal terms in $\overline{\mathbf{H}}[k]$ come from the no-CP property of QAM-FBMC, resulting in inter-symbol interference (ISI). However, this non-diagonal matrix can be approximated by setting the non-diagonal elements as zero so that one-tap equalization in frequency can be performed as

$$\mathbf{G}_{ZF} \triangleq \left( \widehat{\mathbf{H}}^H[0] \widehat{\mathbf{H}}[0] \right)^{-1} \widehat{\mathbf{H}}[0], \quad (6)$$

and

$$\mathbf{G}_{MMSE} \triangleq \left( \widehat{\mathbf{H}}^H[0] \widehat{\mathbf{H}}[0] + \sigma_{I+N}^2 \mathbf{I} \right)^{-1} \widehat{\mathbf{H}}[0], \quad (7)$$

for zero forcing (ZF) and minimum mean square error (MMSE) equalizers, respectively, where $\widehat{\mathbf{H}}[0]$ is a diagonal matrix given by the $N$-point estimated channel, and $\sigma_{I+N}^2$ is the variance of interference plus noise. Then, the filtered signal per each subcarrier after equalization is given by

$$\widetilde{D}_m[0] = \underline{\mathbf{P}}_{f,m}^H \mathbf{G} \underline{\mathbf{X}}_R[0]. \quad (8)$$

where $m$ is the subcarrier index.

### C. Impairment Consideration

Firstly, channel estimation is not always accurate and if the estimation error is taken into consideration, the equalization matrix can be expressed as

$$\mathbf{G}_{ZF} = \left( \left( \widehat{\mathbf{H}}[0] + \widehat{\mathbf{H}}_{er}[0] \right)^H \left( \widehat{\mathbf{H}}[0] + \widehat{\mathbf{H}}_{er}[0] \right) \right)^{-1} \left( \widehat{\mathbf{H}}[0] + \widehat{\mathbf{H}}_{er}[0] \right), (9)$$

and

$$\mathbf{G}_{mmse} = \frac{\widehat{\mathbf{H}}[0] + \widehat{\mathbf{H}}_{er}[0]}{\left( \widehat{\mathbf{H}}[0] + \widehat{\mathbf{H}}_{er}[0] \right)^H \left( \widehat{\mathbf{H}}[0] + \widehat{\mathbf{H}}_{er}[0] \right) + \sigma_{I+N}^2 \mathbf{I}}, \quad (10)$$

respectively, where $\widehat{\mathbf{H}}_{er}$ is the diagonal channel estimation error matrix and its elements are assumed to follow circularly-symmetric complex Gaussian distribution with mean 0 and variance $\sigma_{er}^2$.

Secondly, when phase noise is taken into consideration, $\overline{\mathbf{H}}[k]$ should be rewritten as

$$\overline{\mathbf{H}_{pn}}[k] = \mathbf{W}_N \mathbf{P}_{pn} \mathbf{T}[k] \mathbf{H}[k] \mathbf{W}_N^H \quad (11)$$

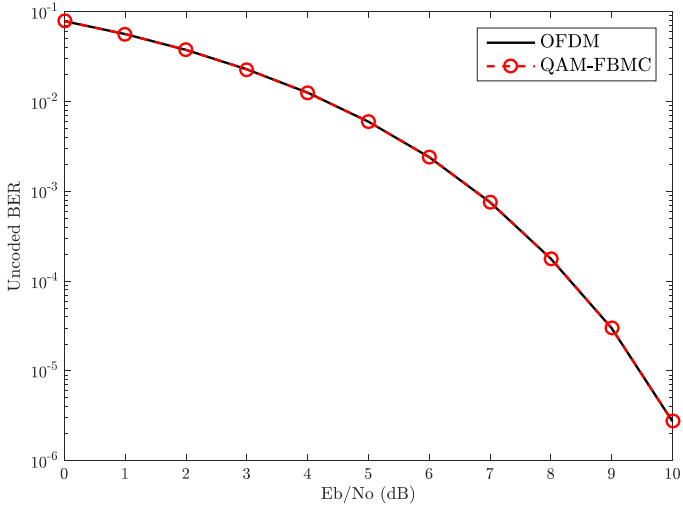

Fig. 2. Uncoded BER in AWGN channel

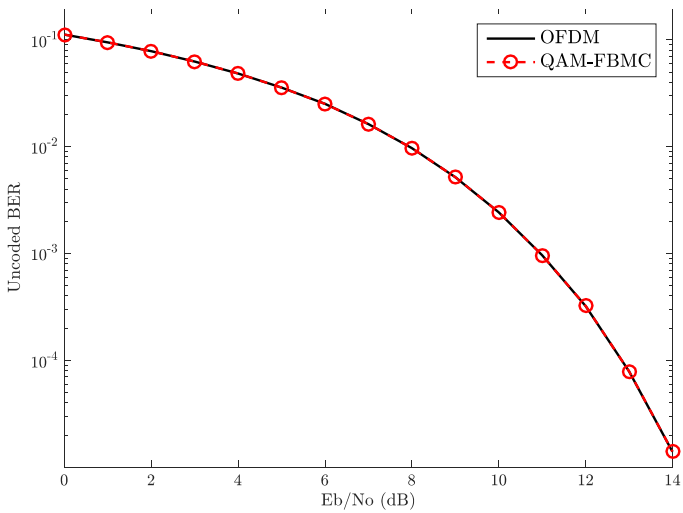

Fig. 3. Uncoded BER in multi-path channel

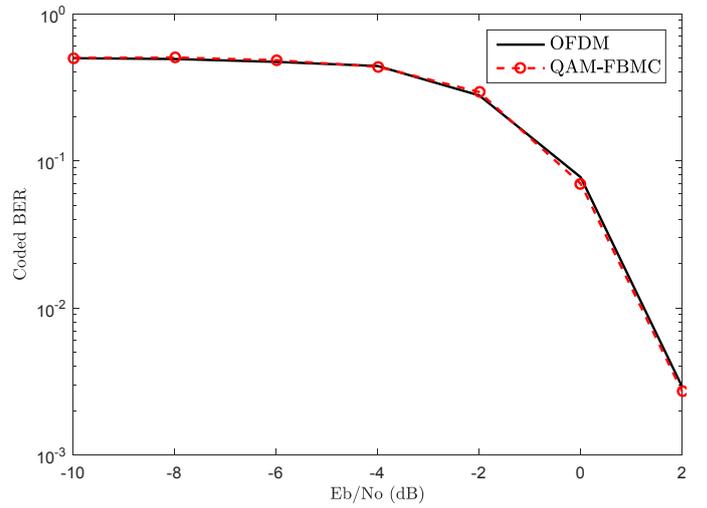

Fig. 4. Coded BER in multi-path channel

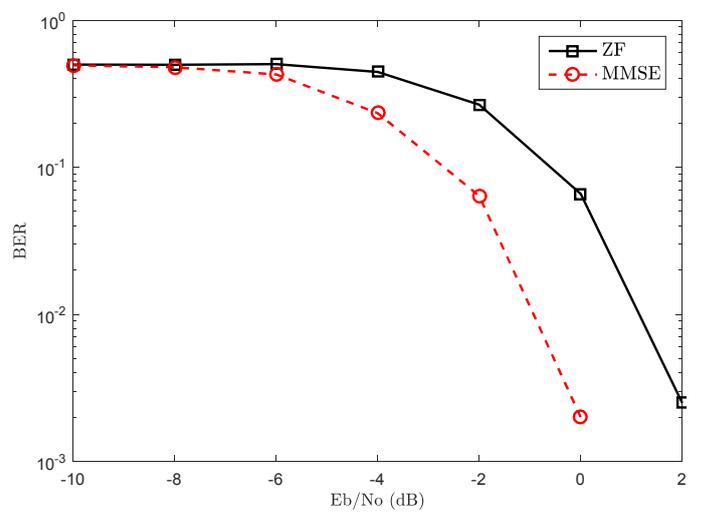

Fig. 5. Perfromance comparison of ZF and MMSE equalization schemes.

where $\boldsymbol{P}_{pn}$ is a diagonal matrix depicting phase noise and its diagonal element is given as

$$\theta[k] := e^{j\phi[k]} \quad (12)$$

We assume that $\theta[k] \approx 1 + j\phi[k]$ for simplicity (valid assumption if $\phi[k] \ll 1$) and eq. (11) can be expressed as

$$\overline{\mathbf{H}_{pn}}[k] = \mathbf{W}_N \mathbf{T}[k]\mathbf{H}[k]\mathbf{W}_N^H + \mathbf{W}_N \boldsymbol{\Phi}_{pn}\mathbf{T}[k]\mathbf{H}[k]\mathbf{W}_N^H, \quad (13)$$

where $\boldsymbol{\Phi}_{pn}$ is a diagonal matrix with diagonal element $j\phi[k]$. It follows that

$$\underline{\mathbf{X}}_R[0] = \sum_{k=-L}^{L-1} \overline{\mathbf{H}}[k]\mathbf{P}_T\underline{\mathbf{D}}[k] + \sum_{k=-L}^{L-1} \overline{\boldsymbol{\Phi}}[k]\mathbf{P}_T\underline{\mathbf{D}}[k] + \underline{\mathbf{w}}[0], \quad (14)$$

where the second term can be treated as additional noise. Then similar operation can be conducted as in eq. (4) to (8).

### D. Complexity, OOB and MIMO Compatibility Evaluations

As aforementioned, the QAM-FBMC system can be efficiently implemented using M/B-point inverse fast Fourier transform (IFFT), *BL*-times repetition, and time domain filtering. Compared with OFDM, the complexity of IFFT is actually reduced from O($M\log_2(M)$) to O($M\log_2(M/2)$) due to the reduced IFFT length. The repetition will only add negligible complexity. The time domain filtering is implemented by element-wise multiplication. At the receiver side, polyphase network (PPN)-FFT scheme can be used with minor complexity addition [7].

With complex domain orthogonally, the conventional MIMO transmission schemes can be utilized with the proposed the QAM-FBMC system [10].

The OOB emissions of QAM-FBMC depend on the design of prototype filters. It has been pointed out in [8] that the OOB emissions can be much lower than OFDM and close to or even the same as OQAM-FBMC. Further evaluation on the effects of phase noise will be presented in next section.

## IV. EVALUATION RESULTS

In this section, simulations are conducted to evaluate the performance of QAM-FBMC in comparison to CP-OFDM. Here, we assume that $M = 1000$, QPSK modulation and $K = 4$.

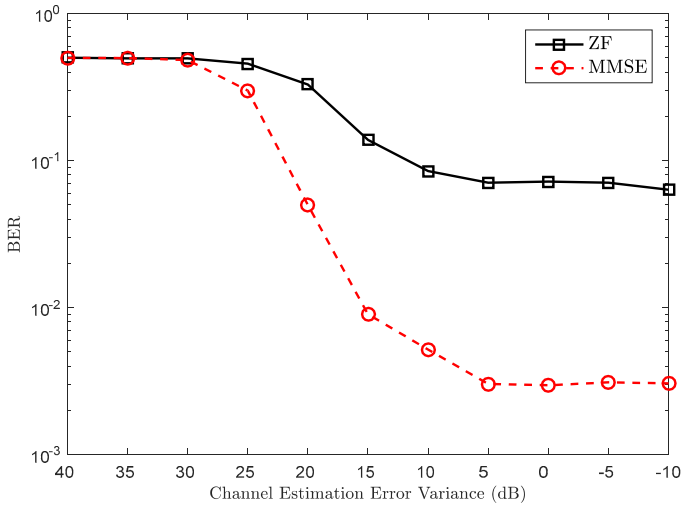

Fig. 6. Effect of channel estimation error ($E_b/N_0 = 0$ dB)

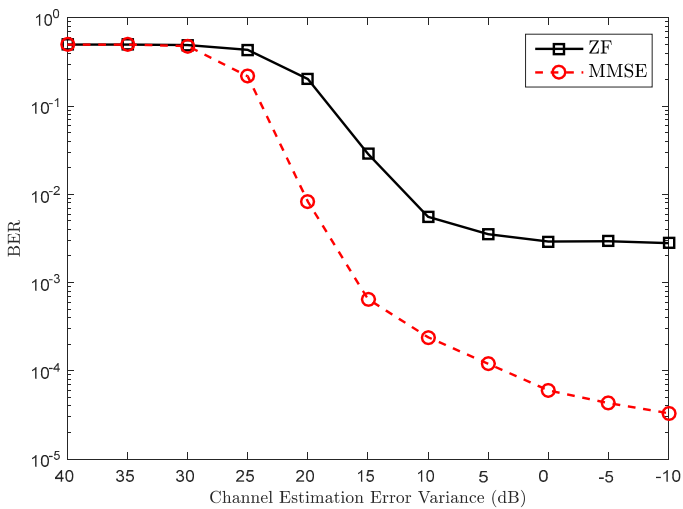

Fig. 7. Effect of channel estimation error ($E_b/N_0 = 2$ dB)

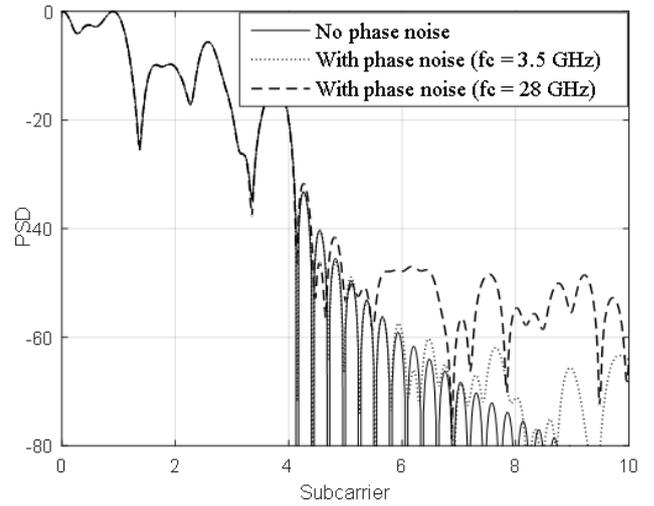

Fig. 8. OOB emission with phase noise (fc: carrier frequency)

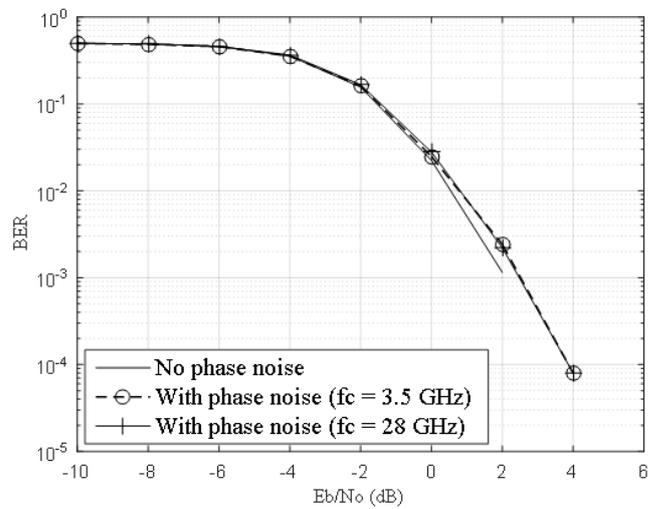

Fig. 9. Effect of phase noise (fc: carrier frequency)

*A. Comparison between CP-OFDM and QAM-FBMC*

The uncoded BER performance is shown in Fig. 2 and Fig. 3 for the AWGN and Extended Vehicular A (EVA) channels, respectively. As it can be seen, the performance of QAM-FBMC is almost the same as CP-OFDM and thus has the same level of robustness against multi-path fading. When convolutional code with coding rate 1/2 is applied as shown in Fig. 4, the BER is improved but the performance of QAM-FBMC and CP-OFDM is still the same.

*B. Comparison of Equalization Schemes and Effects of Channel Estimation Error*

Two different equalization schemes are compared and evaluated as shown in Fig. 5. It is shown that MMSE equalizer outperforms ZF equalizer. However, it should be noted that in MMSE equalizer we assume that the noise variance is known to the receiver for simplicity. We also evaluate the effect of channel estimation error, which is modelled as complex circularly-symmetric Gaussian distribution random variables. Fig. 6 and Fig. 7 show the BER performance with increased channel estimation accuracy for 0 and 2 dB $E_b/N_0$, respectively.

The BER performance is significantly affected by channel estimation error for both AWGN channel and EVA channel. Given the same channel estimation accuracy, MMSE equalizer outperforms ZF but needs more channel estimation accuracy to reach the optimal performance.

*C. Phase Noise*

The high phase noise model in [11] is used in this work to evaluate the robustness of QAM-FBMC. Fig. 8 shows the OOB emission of QAM-FBMC with phase noise. Here type-II filter is assumed [8]. Two frequency points are examined, namely 3.5 GHz and 28 GHz. It is shown that the OOB emission of QAM-FBMC remains in a very low level with phase noise (below -40 dB). Fig. 9 shows the BER with phase noise and it can be seen that the BER performance is only slightly degraded. In short, QAM-FBMC shows significant robustness to phase noise.

V. CONCLUSIONS AND FUTURE WORKS

In this paper, we investigate a newly proposed waveform for 5G – QAM-FBMC, where QAM modulation can be enabled by using multiple prototype filters for different subcarriers. Main challenges and requirements for 5G are identified and waveform candidates are compared with each other. Initial analysis and evaluations are made to the new waveform in a variety of perspectives, e.g., equalization schemes, channel estimation error, effects of phase noise, OOB emission, complexity, etc. Simulation results show that QAM-FBMC can achieve the same performance as CP-OFDM without spectral efficiency reduction. The effects of channel equalizer and channel estimation error are also investigated. Further results show that QAM-FBMC is robust to phase noise. In the future, we will also investigate the reference signalling overhead for QAM-FBMC and impacts of other RF impairments, e.g., power amplifier non-linearity.

ACKNOWLEDGEMENTS

This work has been performed in the framework of the H2020 projects METIS-II and FANTASTIC 5G, co-funded by the EU. The views expressed are those of the authors and do not necessarily represent the projects. The consortia are not liable for any use that may be made of any of the information contained therein.